\documentclass{article}
\usepackage{ijcai25}

\usepackage{times}
\usepackage{soul}
\usepackage{url}
\usepackage[hidelinks]{hyperref}
\usepackage[utf8]{inputenc}
\usepackage[small]{caption}
\usepackage{graphicx}
\usepackage{amsmath}
\usepackage{amsthm}
\usepackage{booktabs}
\usepackage{algorithm}
\usepackage{algorithmic}
\usepackage[switch]{lineno}
\usepackage{booktabs}  
\usepackage{graphicx} 
\usepackage{tabularx}  
\usepackage{subcaption}
\usepackage{xcolor}
\usepackage{mdframed}

\title{Towards Automated Situation Awareness: 
\\A RAG-Based Framework for Peacebuilding Reports}

\author{
Poli A. Nemkova$^{1,2}$
\and
Suleyman O. Polat$^1$\and
Rafid I. Jahan$^1$\and
Sagnik Ray Choudhury$^1$\and
Sun-joo Lee$^2$\and
Shouryadipta Sarkar$^2$\And
Mark V. Albert$^1$\\
\affiliations
$^1$University of North Texas\\
$^2$United Nations Development Programme\\
\emails
\{poli.nemkova, mark.albert, sagnik.raychoudhury\}@unt.edu,
\{suleymanolcaypolat, rafidishrakjahan\}@my.unt.edu,
\{sun-joo.lee, shouryadipta.sarkar\}@undp.org
}

\begin{document}

\maketitle

\begin{abstract}
Timely and accurate situation awareness is vital for decision-making in humanitarian response, conflict monitoring, and early warning and early action. However, the manual analysis of vast and heterogeneous data sources often results in delays, limiting the effectiveness of interventions. This paper introduces a dynamic Retrieval-Augmented Generation (RAG) system that autonomously generates situation awareness reports by integrating real-time data from diverse sources, including news articles, conflict event databases, and economic indicators. Our system constructs query-specific knowledge bases on demand, ensuring timely, relevant, and accurate insights.

To ensure the quality of generated reports, we propose a three-level evaluation framework that combines semantic similarity metrics, factual consistency checks, and expert feedback. The first level employs automated NLP metrics to assess coherence and factual accuracy. The second level involves human expert evaluation to verify the relevance and completeness of the reports. The third level utilizes LLM-as-a-Judge, where large language models provide an additional layer of assessment to ensure robustness. The system is tested across multiple real-world scenarios, demonstrating its effectiveness in producing coherent, insightful, and actionable reports. By automating report generation, our approach reduces the burden on human analysts and accelerates decision-making processes. To promote reproducibility and further research, we openly share our code and evaluation tools with the community via GitHub.
\end{abstract}

\section{Introduction}
Timely situation awareness is essential for effective decision-making in peacekeeping operations, humanitarian response, and governmental interventions. Organizations such as the United Nations (UN), non-governmental organizations (NGOs), and policy-makers rely on accurate and up-to-date reports to allocate resources, anticipate crises, and mitigate conflict escalation. However, generating these reports manually is time-intensive, requiring extensive data collection, synthesis, and expert analysis. The complexity of integrating heterogeneous, real-time data from multiple sources further exacerbates this challenge, leading to delays that can hinder timely interventions.

Advancements in large language models (LLMs) have demonstrated strong capabilities in text aggregation and summarization. However, standard LLM-based approaches face limitations such as hallucination, inconsistencies, and limited fact verification, making them unreliable for high-stakes decision-making. Retrieval-Augmented Generation (RAG) offers a promising alternative by integrating retrieval-based fact grounding with generative models. By dynamically retrieving and incorporating relevant, real-world data into the response generation process, RAG significantly improves factual consistency while reducing hallucinations.

In this paper, we propose a multi-modal dynamic RAG framework designed for automated situation awareness reporting. Our system ingests real-time data from diverse sources, including news articles, conflict event databases, and economic indicators, constructing query-specific knowledge bases to ensure contextually relevant and temporally prompt insights. Reports generated through this system provide stakeholders with structured, evidence-backed summaries, reducing the time and effort required for manual analysis.

To assess the reliability and effectiveness of our system, we introduce a three-level evaluation framework that includes:

Automated NLP-Based Metrics – Assessing factual accuracy, completeness, and coherence using semantic similarity models, factual consistency checks, and bias detection tools.
Human Expert Review – Evaluating relevance, completeness, and usability through domain experts from the UN.
LLM-as-a-Judge Assessment – Benchmarking the model’s performance using LLM-based evaluation, comparing outputs with human judgments.
We evaluate our approach across multiple real-world scenarios and demonstrate that our RAG system produces coherent, insightful, and actionable reports while reducing the burden on human analysts. By automating the early-stage report generation process, we accelerate decision-making while maintaining expert oversight. To encourage further research and reproducibility, we openly share our code, evaluation framework, and sample reports with the community.

To the best of our knowledge, this study presents the first application of a Retrieval-Augmented Generation (RAG) framework integrated with large language models (LLMs) in the peacebuilding domain. Furthermore, it constitutes the first documented instance of AI-driven automation leveraging publicly available data to support decision-making processes for high-impact non-governmental organizations (NGOs), including the United Nations, a key partner in this project.

As a technical contribution, we introduce a dynamic RAG implementation coupled with a three-level reference-free evaluation framework. Additionally, we developed a modified version of VERISCORE \cite{song2024veriscore}, packaged as \textit{ragve}, which enables verification that the RAG-generated outputs are grounded solely in the retrieved data—a valuable tool for researchers working on similar models. The \textit{ragve} package, along with our full implementation, is publicly available on GitHub. 

\section{Literature Review}

Retrieval-Augmented Generation (RAG) represents an advanced methodology that integrates the generative power of large language models (LLMs) with the precision of information retrieval systems. By leveraging external knowledge sources, RAG frameworks enhance the contextual relevance and factual accuracy of generated content. This hybrid approach has demonstrated substantial benefits across diverse fields such as safety, finance, healthcare, and scientific research, significantly improving the quality and efficiency of automated reporting \cite{gao2023retrieval,fan2024survey,arslan2024survey}.

In \textit{safety-critical domains}, RAG frameworks have been effectively customized to generate comprehensive reports based on session logs and incident descriptions. For instance, in the aviation sector, models like LLaMA combined with embedding techniques have automated safety report generation, achieving significant improvements in accuracy and compliance with documentation standards. Metrics such as Recall@5, GLEU, METEOR, and BERTScore highlight the superiority of RAG-based methods over traditional reporting systems \cite{Bernardi2024Automatic,Suresh2024Towards}.

In the \textit{financial sector}, RAG has enhanced the analysis and interpretation of complex financial reports, particularly in question-answering tasks for private investors. Utilizing models like OpenAI’s ADA and GPT-4, these systems process half-yearly and quarterly reports with high contextual relevance and accuracy. Research underscores that well-structured financial documents further optimize RAG performance, especially in addressing qualitative queries \cite{Iaroshev2024Evaluating}.

Similarly, in \textit{healthcare}, RAG has been employed to automate the generation of radiology reports through multimodal embeddings, facilitating the retrieval of relevant clinical information. Generative models, such as OpenAI's GPT series, integrate user-specific clinical requirements to improve report quality while mitigating hallucination risks. These systems have achieved enhanced performance on clinical metrics like BERTScore and Semb score, showcasing the framework's utility in high-stakes environments \cite{Ranjit2023Retrieval,markey2024rags,Wang2024Optimizing,Assistant2024A,alam2024towards}.


While RAG has proven successful across these domains, its application in \textit{peacebuilding} remains an emerging area of exploration. Recent advances in Natural Language Processing (NLP) have demonstrated the potential of LLMs and RAG frameworks for conflict prediction and monitoring human rights violations. Studies have successfully applied NLP techniques to detect armed conflicts and human rights abuses using diverse data sources, including social media and news reports \cite{Trivedi2020How,Alhelbawy2020An,Mueller2024Introducing,Nemkova2023Detecting}. However, comprehensive, automated situation awareness reporting for conflict prevention, resolution, and post-conflict recovery is still underdeveloped. Our work seeks to bridge this gap by extending the proven capabilities of RAG frameworks to peacebuilding, offering a robust solution for humanitarian and governmental organizations.

The evolution of \textit{multimodal RAG systems} has further expanded report generation capabilities by incorporating diverse data types, including text, tables, and images. These systems enhance retrieval and content generation by leveraging interrelationships between different modalities, thereby improving the contextual richness and accuracy of generated reports \cite{Joshi2024Robust,Xia2024MMed-RAG}.


RAG frameworks have demonstrated considerable potential across multiple fields, improving the accuracy, contextual relevance, and efficiency of automated reporting. Our work builds upon these successes by adapting RAG for peacebuilding applications, addressing the unique challenges of conflict monitoring and humanitarian reporting. By doing so, we aim to contribute a novel, practical tool that enhances decision-making in complex geopolitical contexts.

\section{Method and System Design}

\begin{figure*}[!t]
    \centering
    \includegraphics[width=0.7\textwidth]{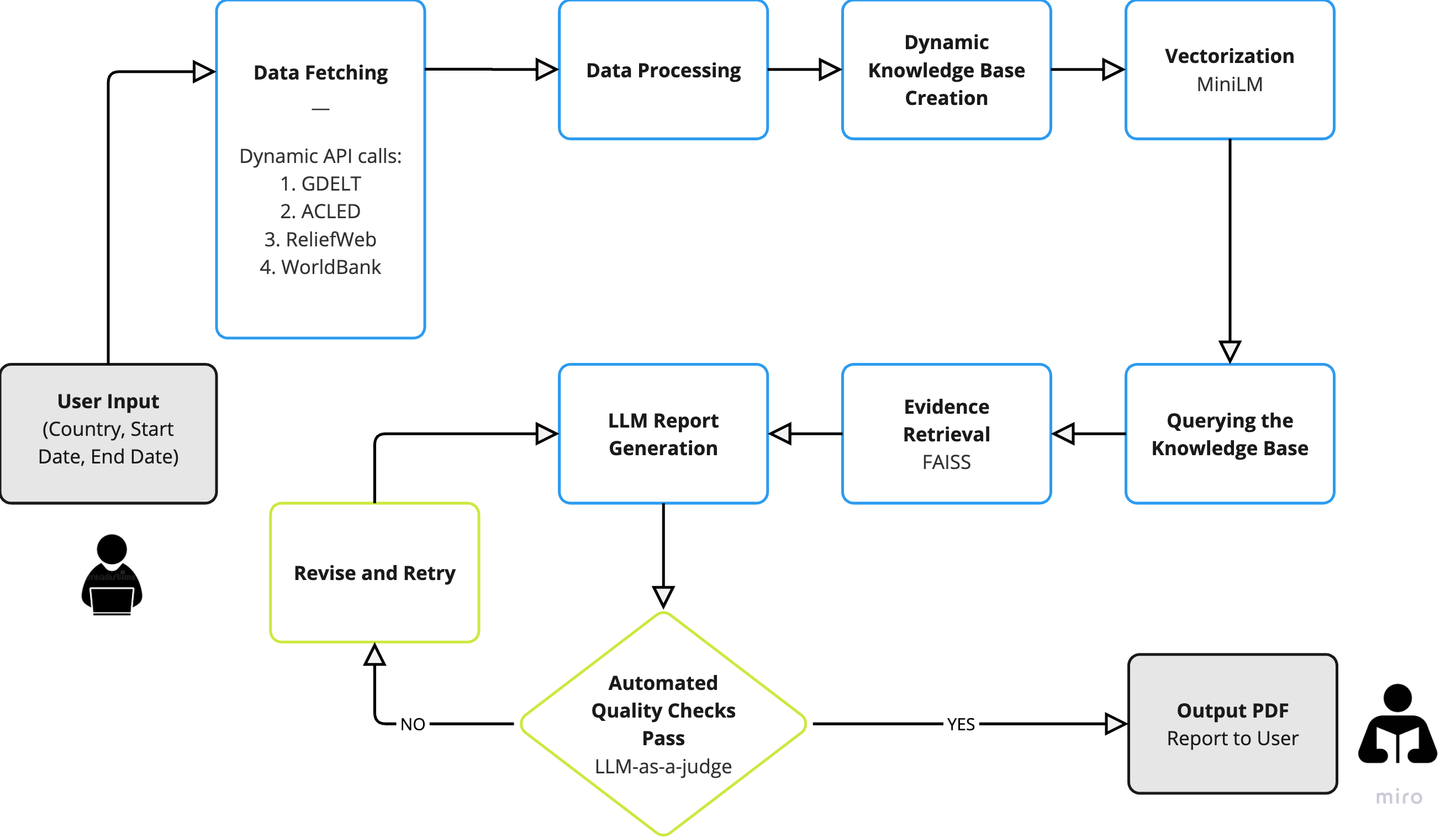}
    \caption{Overall system design presented as a flowchart. Created with Miro.}
    \label{fig:design}
\end{figure*}

To generate comprehensive situation awareness reports, we implemented a dynamic Retrieval-Augmented Generation (RAG) framework. System design can be seen on Figure \ref{fig:design}.

\textit{Data Fetching. }The system operates based on user input specifying the country of interest and the start and end dates of the desired reporting period. Upon receiving these parameters, the system issues four distinct API queries to retrieve relevant data from GDELT, ACLED, ReliefWeb, and the World Bank.

\begin{itemize}
    \item GDELT (Global Database of Events, Language, and Tone)\footnote{https://www.gdeltproject.org/}: GDELT is a publicly available database that captures global political and international affairs events extracted from online news articles. For each relevant event, GDELT provides a link to the original article. After retrieving GDELT event data, we scrape the full text of the referenced news articles to enrich the dataset using library Python \textit{newspaper}\footnote{https://pypi.org/project/newspaper3k/}.
    
    \item ACLED (Armed Conflict Location \& Event Data Project)\footnote{https://acleddata.com/}: ACLED provides structured data on political violence and conflict-related incidents, including numerical indicators such as the number of fatalities \cite{Raleigh2010Introducing}.
    
    \item ReliefWeb\footnote{https://reliefweb.int/}: ReliefWeb, operated by the United Nations Office for the Coordination of Humanitarian Affairs (OCHA), provides blog-style reports and humanitarian briefings on global crises. These articles are retrieved to supplement situational insights.
    
    \item World Bank\footnote{https://pypi.org/project/wbgapi/}: The World Bank API is used to collect relevant economic indicators, including GDP (current US\$), GDP growth (annual \%), inflation rate, unemployment rate, and military expenditure (\% of GDP).
\end{itemize}

\textit{Data Processing. } To ensure compatibility with large language model (LLM) processing, all numerical data from ACLED and the World Bank were converted into textual descriptions by wrapping numbers into a pre-defined textual template. We removed missing values and duplicates for textual data (GDELT and ReliefWeb).

\textit{Vectorization} Following data collection, we employed MiniLM\footnote{https://huggingface.co/sentence-transformers/all-MiniLM-L6-v2} to encode all extracted textual data into vector representations. MiniLM is efficient for encoding as it offers a compact, lightweight architecture that delivers good performance in generating contextualized embeddings while maintaining fast processing speeds and low resource consumption. At this step we have dynamic knowledge base ready.

\textit{Querying the Knowledge Base.} Once the encoded dataset is prepared, we queried the KB with \textit{"Conflict and social unrest issues in \{country\}"}. 

\textit{Evidence Retrieval.} A similarity search, using FAISS (Facebook AI Similarity Search) \cite{douze2024faiss},  was then conducted to retrieve the most relevant data points for each query. Once the top 10 most relevant vector entries were identified, the corresponding full-text content was retrieved. This refined dataset serves as evidence set, which is subsequently used in the report generation process.

 \textit{LLM Report Generation.} After retrieving the most relevant evidence, we used a prompt-based approach to generate a situation awareness report using a Large Language Model (LLM). The prompt is designed to structure the output in a clear and actionable format.
We utilized two prompting strategies: (1) instruction and (2) personification.

The instruction prompt (1) included the following:

   \begin{mdframed}[backgroundcolor=green!8,rightline=false,leftline=false]
Provide a structured summary including the following sections:
\begin{itemize}
    \item Important ongoing situation (if any, optional)
    \item Key recent insights
    \item Trends
    \item Recommendations (label this section as: Recommendation [experimental])
\end{itemize}
Whenever you reference a numerical value or factual information, cite the exact source from which it was obtained in parentheses. Below is the relevant evidence: \{extracted\_earlier\_evidence\}.
\end{mdframed}

This structured prompt ensures that the generated report maintains coherence while providing evidence-based insights. By explicitly instructing the model to cite sources, we enhance the transparency and reliability of the generated content.

The personification prompt (2) included the same strict sectioning, but started with \textit{"You are a conflict analyst preparing a situation awareness report for humanitarian decision-makers. Use the evidence below to craft a clear, concise, and professional report.}"

To evaluate the effectiveness of our approach, we tested our prompt strategies using two different LLMs: GPT-4o\footnote{https://openai.com/index/hello-gpt-4o/} and LLaMA 3\footnote{https://huggingface.co/meta-llama/Meta-Llama-3-8B-Instruct
}. Additionally, we conducted preliminary experiments with DeepSeek \cite{guo2025deepseek}, but the generated outputs were of insufficient quality for our use case. Consequently, we focused our analysis on GPT-4o and LLaMA 3.

The generated reports were saved in both \texttt{.txt} and \texttt{.pdf} formats to facilitate further analysis and sharing. The \texttt{.txt} format was used for structured text processing and comparison, while the \texttt{.pdf} format ensured preservation of formatting and readability across different platforms.

All processing for this version of the model was conducted using Google Colab\footnote{https://colab.research.google.com/} with A100 GPU instances (with Pro+ membership). 

\section{Evaluation}
\textit{Test Sample Selection.} To evaluate the performance of our system, we generated 15 distinct input sets, each comprising a combination of a country, a start date, and an end date. As previously outlined, we employed two different LLM models and applied two distinct prompts. This experimental design resulted in a total of 60 reports for subsequent evaluation\footnote{Sample of the reports can be found on author's Github}. 

       
      
       
       

\begin{table*}[h]
    \centering
    \begin{tabularx}{\textwidth}{@{}l*{4}{>{\centering\arraybackslash}X}@{}}  
        \toprule
        Evaluation Metric & GPT-generated, prompt\_1 & GPT-generated, prompt\_2 & LLaMA-generated, prompt\_1 & LLaMA-generated, prompt\_2 \\
        \midrule
        \multicolumn{5}{@{}l}{\textit{Accuracy}} \\
        Average VERISCORE & 0.76 & 0.73 & 0.91 & 0.79 \\
        Average RAG VERISCORE & 0.65 & 0.69 & 0.60 & 0.61 \\
        
        \multicolumn{5}{@{}l}{\textit{Consistency}} \\
        SummaC & 0.52 & 0.57 & 0.64 & 0.67 \\
        
        \multicolumn{5}{@{}l}{\textit{Objectivity/Bias}} \\
        Politically Center Confidence & 0.99 & 0.99 & 0.99 & 0.99 \\
        
        \multicolumn{5}{@{}l}{\textit{Coherence/Clarity}} \\
        Coherence Score & 0.79 & 0.81 & 0.81 & 0.84 \\
        \bottomrule
    \end{tabularx}
    \caption{Automated Evaluation of Reports with NLP methods (Level 1 Evaluation)}
    \label{table:nlp_eval}
\end{table*}
To ensure a comprehensive evaluation, we generated reports for diverse geographical regions, covering a wide range of geopolitical and conflict dynamics. The selected countries included:

\begin{table*}[h]
    \centering
    \begin{tabularx}{\textwidth}{@{}l*{3}{>{\centering\arraybackslash}X}@{}}  
        \toprule
        Evaluation Metric & GPT-as-a-judge & LLaMA-as-a-judge & Claude-as-a-judge \\
        \midrule
        \\
        Average Score for GPT generated reports & 1 & 0.82 & 0.95 \\
        Average Score for LLaMA generated reports & 0.93 & 0.78 & 0.98 \\
        Poorly Performed Questions  with GPT generated reports & None & Q3 & Q3\\
        Poorly Performed Questions  with LLaMA generated reports & Q4 &  Q3 & Q3\\
        \\
        \bottomrule
    \end{tabularx}
    \caption{Comparison of GPT and LLaMA judgment of the reports (Level 3 Evaluation)}
    \label{table:llm_judge}
\end{table*}

\begin{itemize}
    \item Middle East (ME): Iran, Israel, Syria, Lebanon, Yemen
    \item Eastern Europe (EE): Ukraine, Russia
    \item Horns of Africa (HOA): Sudan, Ethiopia, Somalia, South Sudan
    \item Asia: Myanmar, China
\end{itemize}

Reports were generated for different timeframes, specifically covering periods of 1 month, 3 months, and 1 year.

\begin{figure}[!t]
    \centering
    \includegraphics[width=0.2\textwidth]{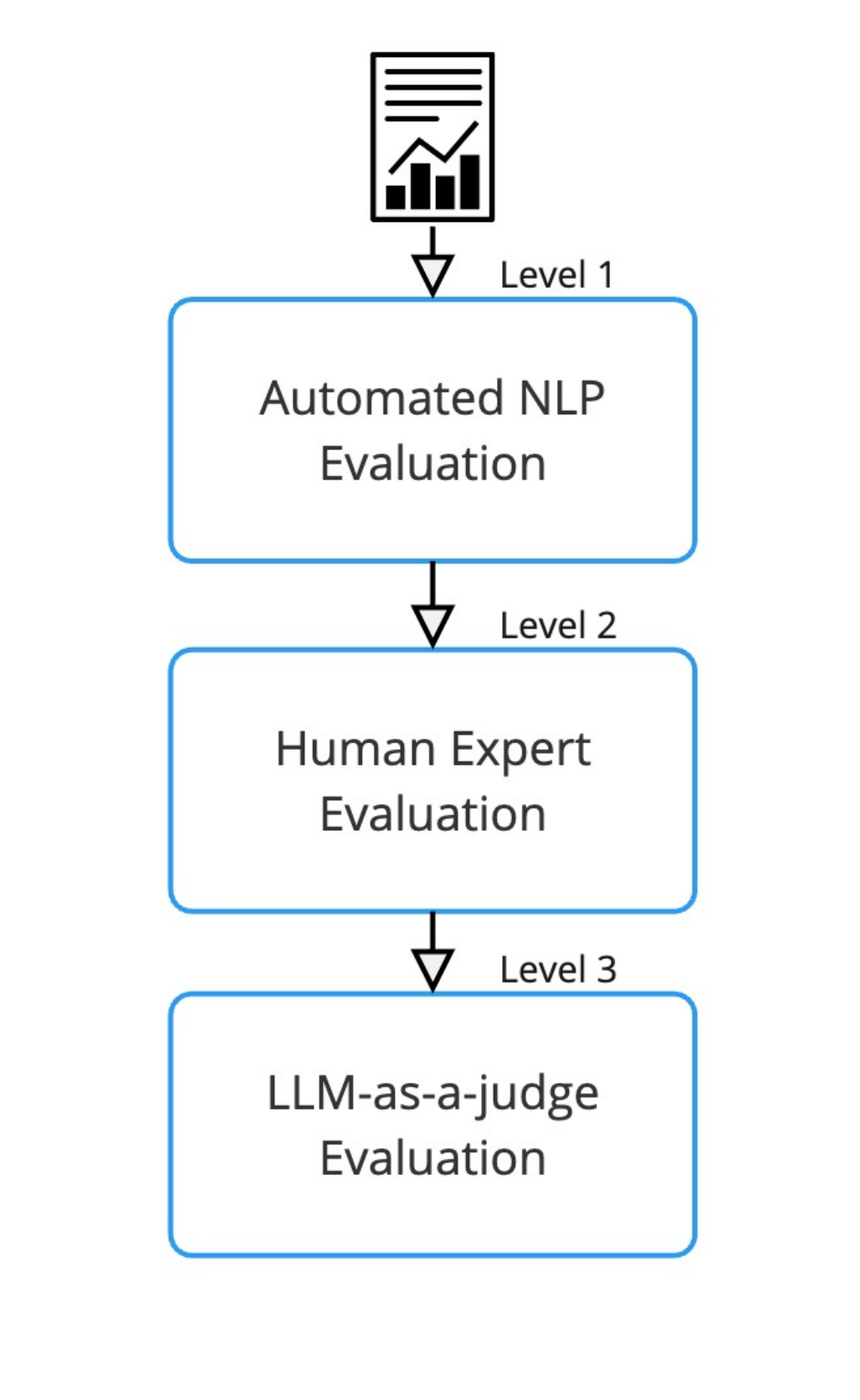}
    \caption{Report evaluation three-level framework. Created with Miro.}
    \label{fig:eval}
\end{figure}

To assess the performance and quality of the generated situation awareness reports, we implemented a three-layer evaluation framework as presented on Figure \ref{fig:eval}.\\

\textbf{Level 1: NLP-Based Automated Metrics}

At the first evaluation level, we leveraged existing NLP models to measure key text quality attributes. The following metrics were used:

\begin{itemize}
    \item \textit{Accuracy:} To assess factual correctness and detect potential hallucinations, we employed VERISCORE tool \cite{song2024veriscore}. We used the tool to check the accuracy against most recent Google search. And also, we modified the tool to check the accuracy against the dynamic knowledge base - we provide a new library for this adjusted tool on GitHub.\footnote{https://github.com/withheld-for-anonymity}
    \item \textit{Consistency:} To evaluate consistency, we assessed whether the generated reports accurately retained the key points from the original, larger text. This task was conceptualized as analogous to summary evaluation, given the focus on preserving essential information. Accordingly, we utilized the SummaC library \cite{laban2022summac}, a tool specifically designed for evaluating the quality and fidelity of summaries.
    \item \textit{Objectivity/Bias:} To identify potential bias in the generated reports, we utilized politicalBiasBERT \cite{baly2020we}, a political classifier, as implemented on HuggingFace\footnote{https://huggingface.co/bucketresearch/politicalBiasBERT}.
    \item \textit{Coherence/Clarity:} To evaluate the structural and linguistic coherence of the reports, we computed a coherence score based on BERT-based text similarity measures \cite{devlin2018bert}.
\end{itemize}

Only reports that met acceptable threshold values in Level 1 evaluation were forwarded to the next stage, Level 2: Human Expert Evaluation. \\

\textbf{Level 2: Human Expert Evaluation}

At this stage, the reports were assessed by two Human Expert Evaluators, who are senior employees at United Nations Development Programme (UNDP)\footnote{https://www.undp.org/} and represent the target users of the generated reports. The evaluation panel consisted of both male and female experts to ensure diverse perspectives in the assessment process. Each report was independently evaluated by two experts, ensuring that each report received two separate evaluation scores.

\textit{Evaluation Criteria}. The human expert evaluators assessed each report based on a structured set of evaluation questions categorized into two main parts: Relevancy and Completeness (binary evaluation) and Preference-Based Comparisons.\\

\textbf{Part A}: \textit{Relevancy and Completeness (True/False)}. 
Each evaluator answered the following binary questions to assess the report's quality:

\begin{itemize}
        \item Q1. Is the report relevant?
        \item Q2. Does more than 50\% of the report contain relevant information?
        \item Q3. Does more than 90\% of the report contain relevant information?
        \item Q4. Does the report avoid duplicate information?
        \item Q5. Does the report contain no more than 10\% of the irrelevant information?
        \item Q6. Does the report seem to be complete?
        \item Q7. Does the report cover economic, political, social, or humanitarian aspects?
\end{itemize}

\textbf{Part B}: \textit{Preference-Based Comparison.}
In addition to the binary evaluation, the experts were asked to compare pairs of reports and indicate their preferences:

\begin{itemize}
    \item Q8. Which report is more complete? (Report 1 vs. Report 2, Report 3 vs. Report 4)
    \item Q9. Which report is more accurate? (Report 1 vs. Report 2, Report 3 vs. Report 4)
    \item Q10. Which report do you prefer overall? (Report 1 vs. Report 2, Report 3 vs. Report 4)
\end{itemize}

This evaluation framework ensures a rigorous assessment of the reports' factual accuracy, completeness, and overall usability from the perspective of expert users.\\

\textbf{Level 3: LLM-as-a-Judge Evaluation.}

To ensure the potential scalability of our evaluation framework, we implemented an LLM-as-a-Judge approach. Specifically, we prompted GPT-4o and LLaMA 3 to evaluate all reports using the same questionnaire as the Human Expert Evaluators. To assess and avoid model's self bias, we also used third LLM - Claude 2\footnote{https://www.anthropic.com/claude} to evaluate the reports.

\section{Results and Discussion}
\textbf{Human evaluation}. Results of human evaluation are presented in Table \ref{table:human_eval}. The inter-annotator agreement between the two human experts, measured using Cohen's Kappa \cite{cohen1960coefficient}, indicated a moderate level of agreement, with values of 0.54 for GPT-generated reports and 0.57 for LLaMa-generated reports.

In the binary evaluation of Part A (Relevance and Completeness), GPT-generated reports achieved an average of 62\% of the total possible points, while LLaMa-generated reports scored slightly higher at 64\%.

In the preference-based evaluation of Part B, human experts selected GPT-generated reports in 76\% of cases, whereas LLaMA-generated reports were preferred in only 24\% of cases.

Notably, the aspects that consistently received the lowest scores were largely similar for both GPT-generated and LLaMa-generated reports. These included questions 4, 5, and 7, which focus on issues of redundant information and the omission of specific aspect of coverage. Additionally, LLaMa-generated reports demonstrated particular difficulties with question 6, which pertains to the completeness of the reports.

\textbf{LLM-as-a-Judge.} Notably, GPT assigned an average perfect score of 100\% to reports generated by itself, compared to 93\% for those produced by LLaMA. The primary weakness identified by GPT across both sets of reports was the presence of redundant information, as reflected in Question 4 (Q4) of the evaluation criteria. This issue was more pronounced in the LLaMA-generated reports, which received the lowest scores in this category.

\begin{table*}[h]
    \centering
    \begin{tabularx}{\textwidth}{@{}l*{4}{>{\centering\arraybackslash}X}@{}}  
        \toprule
        Evaluation Metric & GPT-generated, prompt\_1 & GPT-generated, prompt\_2 & LLaMA-generated, prompt\_1 & LLaMA-generated, prompt\_2 \\
        \midrule

        Cohen's Kappa Overall & 0.54 & \textbf{0.57} & 0.42 & 0.54 \\

        \multicolumn{5}{@{}l}{\textit{Binary Evaluation}} \\
        Cohen's Kappa on Binary Evaluation & 0.53 & 0.51 & 0.50 & \textbf{0.61} \\

        \multicolumn{5}{@{}l}{\textit{Preference-based Evaluation}} \\
        Cohen's Kappa on Preference Evaluation & \textbf{0.52} & \textbf{0.52} & 0.26 & 0.31 \\
        Cohen's Kappa for Q8: Which report is more complete? & \textbf{0.58} & \textbf{0.58} & 0.24 & 0.12 \\
        Cohen's Kappa for Q9: Which report is more accurate? & \textbf{0.54} & \textbf{0.54} & 0.17 & 0.12 \\
        Cohen's Kappa for Q10: Which report do you prefer overall? & 0.44 & \textbf{0.54} & 0.36 & 0.48 \\

        Avg. Max Score (Binary Questions) & 0.62 & \textbf{0.64} & 0.60 & 0.63 \\
        Preferred Report (\%) & \textbf{0.76} & 0.24 & 0.65 & 0.35 \\

        Poorly Performed Questions & Q4, Q5, Q7 & Q4, Q5, Q6, Q7 & Q4, Q5 & Q4, Q5 \\
        Regional Best Performance & Asia, EE & HOA, ME & Asia, ME & Asia, EE \\
        Regional Worst Performance & HOA, ME & Asia, EE & EE, HOA & ME \\

        \bottomrule
    \end{tabularx}
    \caption{Comparison of GPT and LLaMA Reports Based on Human Expert Evaluation (Level 2 Evaluation)}
    \label{table:human_eval}
\end{table*}


\subsection{Human vs. LLM Evaluation}
The highest Cohen's Kappa values from humans were around 0.57 (GPT-generated, prompt 2) and 0.54 (LLaMA-generated, prompt2). These values indicate moderate agreement among human evaluators.
For binary evaluations, humans showed varying agreement, with LLaMA-generated reports (prompt 2) achieving the highest human agreement (0.61).
GPT-as-a-judge gave itself a perfect 1.0 score, while Claude scored GPT reports at 0.95—both suggesting near-perfect evaluations.
LLaMA, however, was more critical of both itself (0.78) and GPT (0.82).

Human evaluations show more variability and critical judgment, reflected in moderate Cohen's Kappa scores, while LLM judges—especially GPT and Claude—tend to rate reports significantly higher, hinting at possible overconfidence or evaluation bias in LLMs.
LLaMA-as-a-judge mirrors human evaluators to some extent by being more critical, particularly towards its own reports.
Interestingly, Claude-as-a-judge preferred LLaMA-generated reports (0.98) slightly more than GPT reports (0.95), suggesting Claude's evaluation diverges from human preferences.
GPT-as-a-judge rated its own reports higher than LLaMA's, aligning more closely with human preferences.
While GPT-as-a-judge aligns with human preferences by favoring its own reports, Claude shows a bias toward LLaMA-generated reports, diverging from human evaluators. This indicates that LLM judges may not always reflect human judgment, especially when cross-model evaluations are involved.
Human evaluators and LLM judges identify different weaknesses in the reports. Humans consistently highlight Q4 and Q5, while LLMs focus on Q3. This discrepancy suggests that LLMs and humans prioritize different aspects of report quality or interpret the evaluation criteria differently.

\subsection{Strengths and Limitations of the Approach}

\textbf{Strengths:}
\begin{itemize}
    \item The system dynamically retrieves relevant information, ensuring reports are evidence-based. All the data used in the study is free and publicly available.
    \item The multi-layered evaluation framework enhances reliability and robustness. Human Evaluation is only used for aligning automated evaluation that will be needed for scaling.
    \item The use of multiple LLMs (GPT-4o, LLaMA 3) allows for comparative analysis and improved output quality.
    \item The generated reports significantly reduce the time required for human analysts to draft reports from scratch. Currently a human analyst, on average, can take up to 2 weeks to create similar report. With our system, this time can drop to 1 week (generated report is used as a base for review and refinement).
\end{itemize}

\textbf{Limitations:}
\begin{itemize}
    \item Potential biases in retrieved evidence may affect the objectivity of the reports.
    \item LLMs may struggle with complex geopolitical nuances and context-dependent interpretations.
    \item Human evaluation introduces subjectivity.
    \item Despite automation, human review is still mandatory before reports can be delivered to stakeholders (human-in-the-loop requirement).
\end{itemize}

\subsection{Regional Variations in Model Performance}

In our study, we observed distinct variations in model performance across different regions. While GPT models generally outperformed LLaMA in overall reporting quality, region-specific differences emerged. GPT demonstrated superior performance in generating reports for Asia and Eastern Europe, whereas LLaMA produced more accurate and contextually relevant outputs for the Middle East and Africa.

A notable factor influencing these results is the disparity in media coverage across regions. Events in Europe and the Middle East tend to receive significantly more international attention compared to regions like the Horn of Africa. This uneven distribution of data likely contributes to variations in model performance, as regions with limited coverage may present greater challenges for accurate information retrieval and synthesis.

\subsection{Benefits for Real-World Applications}  

The proposed system offers several distinct advantages for practical implementation:  

1. \textit{Enhanced Time Efficiency}: 
   The manual production of comparable analytical reports typically requires up to two weeks of continuous effort by a human analyst. Our system reduces this time by approximately 50\%, generating a preliminary report that serves as a foundation for further refinement. This significantly accelerates the reporting pipeline while maintaining analytical rigor.  

2. \textit{Scalability and Expanded Coverage}:  
   Human resource constraints often limit the geographical or thematic scope that analysts can feasibly cover. By automating the initial stages of report generation, our system enables broader coverage across multiple regions or topics on a more frequent or regular basis, thereby enhancing the scalability of monitoring and analysis efforts without proportional increases in staffing.  

3. \textit{Resource Optimization and Cost Efficiency: }
   The system exclusively utilizes publicly available, open-access data sources, eliminating the need for costly proprietary datasets. This approach not only reduces operational expenditures but also makes the system particularly suitable for resource-constrained organizations, such as NGOs and humanitarian agencies.  

4. \textit{Transparency and Reproducibility:}  
   By leveraging open data, the system ensures transparency in data sourcing and analytical processes. This facilitates reproducibility and fosters trust among stakeholders, including policymakers, researchers, and civil society actors, who can validate and build upon the generated reports.
   
5. \textit{Rapid Situational Awareness:} 
    In the event of a sudden conflict outbreak, the system can generate automated reports that offer stakeholders an immediate preliminary assessment of the situation. This rapid access to critical information enables timely decision-making and response, bridging the gap before official reports are published.

\section{Future Work}

This version of the system serves as an initial implementation, with planned improvements in the following areas:
\begin{itemize}
    \item Visualizations to improve data representation and readability.
    \item Forecasting sections to provide predictive insights for stakeholders.
    \item Analytical sections that integrates or synthesizes information to detect new trends.
    \item Refining bias detection and mitigation strategies.
    \item Expanding LLM-as-a-judge to other evaluation tasks.
    \item Providing length and style options on the input stage.
\end{itemize}

\section{Conclusion}
In this work, we introduced a multimodal dynamic Retrieval-Augmented Generation (RAG) system for automated situation awareness reporting, designed to support peacekeeping operations, humanitarian organizations, and government agencies in making timely and informed decisions. By integrating real-time data from diverse sources, including news articles, conflict databases, and economic indicators, our system provides comprehensive and actionable reports while reducing the burden on human analysts.

While our automated reporting system offers a significant step forward in streamlining intelligence-gathering processes, it is not intended to replace human expertise but rather to augment it by providing preliminary insights quickly and efficiently. Future work will focus on refining the system's adaptability to new data sources, improving its interpretability, and expanding the evaluation framework to further validate its effectiveness across diverse operational contexts.

To foster further research and collaboration, we openly share our code and evaluation framework, encouraging the development of enhanced AI-driven solutions for dynamic situation awareness reporting.

\section*{Ethical Statement}
All data utilized in this project is sourced exclusively from publicly accessible APIs, ensuring transparency and verifiability.  No proprietary or confidential data is used, maintaining compliance with ethical and legal standards.

\section*{Acknowledgments}

We sincerely appreciate the support of the United Nations Development Programme (UNDP) Crisis Bureau for introducing the research problem and providing valuable insights throughout the study. Their comprehensive feedback and expert evaluation were instrumental in refining our approach and assessing the effectiveness of the generated reports.

\bibliographystyle{named}
\bibliography{ijcai25}

\end{document}